\documentclass[english,twocolumn,showpacs,nofootinbib]{revtex4}
\usepackage[T1]{fontenc}
\usepackage[utf8]{inputenc}
\usepackage{amsmath}
\usepackage{amssymb}
\usepackage{graphicx}
\usepackage{esint}

\makeatletter
\@ifundefined{textcolor}{}
{%
 \definecolor{BLACK}{gray}{0}
 \definecolor{WHITE}{gray}{1}
 \definecolor{RED}{rgb}{1,0,0}
 \definecolor{GREEN}{rgb}{0,1,0}
 \definecolor{BLUE}{rgb}{0,0,1}
 \definecolor{CYAN}{cmyk}{1,0,0,0}
 \definecolor{MAGENTA}{cmyk}{0,1,0,0}
 \definecolor{YELLOW}{cmyk}{0,0,1,0}
 }

\makeatother

\usepackage{babel}
\begin{document}

\title{Traversable Wormholes and Time Machines in non-minimally coupled
curvature-matter $f(R)$ theories}

\author{Orfeu Bertolami}

\altaffiliation{Also at Instituto de Plasmas e Física Nuclear, Instituto Superior Técnico, Av. Rovisco Pais 1, 1049-001 Lisboa, Portugal}

\email{orfeu.bertolami@fc.up.pt}

\selectlanguage{english}%

\author{Ricardo Zambujal Ferreira}

\email{ricardozambujal@gmail.com}

\selectlanguage{english}%

\affiliation{Departamento de Física e Astronomia, Faculdade de Ciências, Universidade
do Porto,\\
Rua do Campo Alegre 687, 4169-007 Porto, Portugal}
\begin{abstract}
We obtain traversable wormhole and time machine solutions of the field
equations of an alternative of gravity with nonminimally curvature-matter
coupling. Our solutions exhibit a nontrivial redshift function and
allow for matter that satisfies the dominant energy condition.
\end{abstract}


\maketitle

\section{INTRODUCTION}

General Relativity (GR) can account, in the context of the cosmological
standard model, for all cosmological observations provided two unknown
constituents, dark energy ($\Omega_{DE}\simeq0.73$) and dark matter
($\Omega_{DM}\simeq0.22$) are considered in the stress-energy tensor
of Einstein field equations. Given that the nature of dark energy
and dark matter is unknown it is quite natural that alternative theories
of gravity are considered alongside with proposals for dark energy
and dark matter (see e.g. Refs. \cite{Bertolami:2006js,Bertolami:2006fh,Bertolami:2011aa}).

In this respect, a particularly interesting alternative to GR is the
broad class of theories arising from replacing the linear dependence
of scalar curvature in the action of GR by a more general function,
the so-called $f(R)$ theories \cite{RevModPhys.82.451}. In the context
of this extension one might also question the assumption that gravity
is coupled nominimally with matter \cite{Bertolami:2007gv,PhysRevD.72.063505}.
A nonminimally coupling between matter and curvature gives rise to
a deviation of the geodesic motion of test particles, nonconservation
of the stress-energy tensor and many other striking features. These
also include the breaking of the degeneracy of the Lagrangian densities
which, in GR, give rise to the stress-energy tensor of the perfect
fluid \cite{Bertolami:2008ab}, deviation from the hydrodynamic equilibrium
of stars \cite{Bertolami:2007vu}, mimicking of dark matter in galaxies
\cite{Bertolami:2009ic} and clusters of galaxies \cite{Bertolami:2011ye},
of dark energy at cosmological scales \cite{Bertolami:2010cw} and
somewhat more natural conditions for preheating in inflationary models
\cite{Bertolami:2010ke}. It is also shown that the nonminimal coupling
between matter and curvature can be interpreted, under conditions,
as an effective pressure leading to a generalization of the Newtonian
gravitational potential in the weak field limit \cite{Bertolami:2011rb},
and to mimic a cosmological constant for a suitable matter distribution
\cite{Bertolami:2011fz}.

In this work we examine the role played by the nonminimal coupling
in wormhole geometries, namely traversable wormholes, and on the possibility
of generating closed timelike curves (CTCs). Wormholes in classical
GR are rather exotic objects. In order to ensure that gravity is attractive
the Raychaudhuri's equation for the expansion of a congruence of geodesics
defined by a tangent vector field $u^{\mu}$ states that $R_{\mu\nu}u^{\mu}u^{\nu}\geq0$,
which, using Einstein's equations, implies that $\left(T_{\mu\nu}-\frac{T}{2}g_{\mu\nu}\right)u^{\mu}u^{\nu}\geq0$.
This last condition is usually referred to as strong energy condition
and it directly implies the null energy condition (NEC), which states
that $T_{\mu\nu}k^{\mu}k^{\nu}\geq0$ where $k^{\alpha}$ is a null
vector. The NEC, if applied for instance to a perfect fluid, implies
that $\rho+p\geq0$. However, in order to have wormhole solutions
it is required the violation of the NEC in a region containing the
wormhole throat \cite{Morris:1988cz}.

On the other hand, there are two other conditions that are verified
by the stress-energy tensor of all known types of matter: the dominant
energy condition (DEC) which implies for a perfect fluid that $\rho>0$
and $p\in\left[-\rho,\rho\right]$, meaning that the sound velocity
cannot exceed the speed of light, and the weak energy condition which
states that $\rho>0$ and $\rho+p>0$. The DEC implies the weak energy
condition and this implies the NEC. Thus, if the NEC is violated the
three other energy conditions are also violated. In GR, this implies
that exotic and unknown forms of matter are needed to obtain wormhole
solutions so that observers perceive negative energy densities.

One of the most striking features of stable wormhole solutions is
that one can generate CTCs from them \cite{Morris:1988tu}. This can
give origin to controversy and one can wonder whether traversable
wormholes can be realistically created \cite{Morris:1988cz}. Given
the above requirements on the energy density and pressure, several
effects of quantum nature have been invoked. For instance, it has
been argued that these exotic behaviors might arise, due to the Casimir
effect, gravitational backreaction and other effects. However, given
that these effects most often lead to instabilities that prevent wormhole
and CTCs, they actually turn impossible any form of time travel (see
Ref. \cite{Visser:1992tx} for a review). As we shall see, CTCs out
of wormhole solutions can be obtained, under conditions, in the context
of nonminimal curvature-matter coupled theories even for ordinary
matter.

The present work extends the results of Refs \cite{MontelongoGarcia:2010xd,Garcia:2010xb},
where exact wormhole solutions were obtained for a trivial redshift
function, the function that defines the $g_{00}$ component of the
metric. As will be seen, theories of gravity with nonminimal matter-curvature
coupling admit solutions that violate the NEC even for ordinary matter
for the most general type of wormholes, and these can give origin
to CTCs.

This paper is organized as follows: Section II presents a brief outline
of the nonminimal curvature-matter coupling in $f(R)$ theories of
gravity. In Sec. III, we introduce the wormhole geometry supported
by this type of modified theories of gravity. We consider the field
equations for a perfect fluid. In Sec. IV, we look for traversable
wormhole solutions in some specific limits. We analyze the violation
of the NEC and we relate it with the possibility of time travel. In
Sec. V, we discuss our results and present our conclusions.

\section{NONMINIMAL CURVATURE-MATTER COUPLING IN $f(R)$ THEORIES}

The action for a nonminimal curvature-matter coupling in $f(R)$ theories
is given by \cite{Bertolami:2007gv}

\begin{equation}
S=\int\left[\frac{1}{2k}f_{1}(R)+\left(1+\lambda f_{2}(R)\right){\cal {\cal L_{M}}}\right]\sqrt{-g}d^{4}x,\label{eq:Nonminimal coupling action}
\end{equation}
where $k^{2}=8\pi G$, $f_{1},f_{2}$ are arbitrary functions of the
scalar curvature, $R$, and ${\cal L_{M}}$ is the matter Lagrangian
density. The coupling constant $\lambda$ characterizes the strength
of the interaction between curvature and matter and has suitable units.
Notice that theories with similar features have also been examined
in the context of late time-accelerating universes \cite{PhysRevD.72.063505}.

Varying the action with respect to the metric we obtain the field
equations{\small{} and adapting that $k^{2}=1$:}{\small \par}

\[
F_{1}(R)R_{\mu\nu}-\frac{1}{2}f_{1}(R)g_{\mu\nu}=\nabla_{\mu}\nabla_{\nu}F_{1}(R)-g_{\mu\nu}\square F_{1}(R)
\]

\begin{equation}
+2\lambda\left(\Delta_{\mu\nu}-R_{\mu\nu}\right){\cal L_{M}}F_{2}(R)+\left(1+\lambda f_{2}(R)\right)T_{\mu\nu}^{(m)},\label{eq:Field equations}
\end{equation}
where $F_{i}\equiv\frac{df_{i}(R)}{dR}$ , $\Delta_{\mu\nu}\equiv\nabla_{\mu}\nabla_{\nu}-g_{\mu\nu}\square$
and $T_{\mu\nu}^{(m)}$ is the usual stress-energy tensor of matter
defined as
\begin{equation}
T_{\mu\nu}^{(m)}=\frac{-2}{\sqrt{-g}}\frac{\delta\left(\sqrt{-g}{\cal L_{M}}\right)}{\delta\left(g^{\mu\nu}\right)}.\label{eq:Normal matter stress-energy tensor definition}
\end{equation}

Equation (\ref{eq:Field equations}) can be rewritten in a more conventional
form in terms of the Einstein's tensor

\begin{equation}
R_{\mu\nu}-\frac{1}{2}Rg_{\mu\nu}\equiv G_{\mu\nu}=T_{\mu\nu}^{eff},\label{eq:Einstein Equation modified}
\end{equation}
where the effective stress-energy tensor has been defined as{\small 
\[
T_{\mu\nu}^{eff}=\frac{1}{F_{1}}\left[\left(\nabla_{\mu}\nabla_{\nu}-g_{\mu\nu}\left(\square+\frac{1}{2}R\right)\right)F_{1}(R)+\frac{1}{2}g_{\mu\nu}f_{1}(R)\right.
\]
}{\small \par}

{\small 
\begin{equation}
\left.+2\lambda\left(\Delta_{\mu\nu}-R_{\mu\nu}\right){\cal L_{M}}F_{2}(R)+\left(1+\lambda f_{2}(R)\right)T_{\mu\nu}^{(m)}\right].\label{eq:Definition of effective tensor}
\end{equation}
}{\small \par}

Applying the Bianchi identity, $\nabla^{\mu}G_{\mu\nu}=0$, in Equation
(\ref{eq:Field equations}) and using the relation
\begin{equation}
\left(\square\nabla_{\nu}-\nabla_{\nu}\square\right)F_{i}=R_{\mu\nu}\nabla^{\mu}F_{i}\label{eq:Useful relation}
\end{equation}
we obtain for the stress-energy tensor of matter
\begin{equation}
\nabla^{\mu}T_{\mu\nu}^{(m)}=\frac{\lambda F_{2}}{1+\lambda f_{2}}\left[g_{\mu\nu}{\cal L_{M}}-T_{\mu\nu}^{(m)}\right]\nabla^{\mu}R,\label{eq:Derivative of the stress energy tensor}
\end{equation}
meaning that its covariant derivative does not vanish automatically.

Eq. (\ref{eq:Derivative of the stress energy tensor}) implies that
the motion of a test particle is nongeodesic as an extra force shows
up \cite{Bertolami:2007gv}
\begin{equation}
\frac{dU^{\mu}}{ds}+\Gamma_{\alpha\beta}^{\mu}U^{\alpha}U^{\beta}=f^{\mu}.\label{eq:Equation of Motion with extra force}
\end{equation}

For the specific case of a perfect fluid with stress-energy tensor
given by 
\begin{equation}
T_{\mu\nu}^{(m)}=(\rho+p)U_{\mu}U_{\mu}+pg_{\mu\nu},\label{eq:Simple perfect fluid stress-energy tensor}
\end{equation}
where $\rho$ is the energy density, $p$ is the pressure and $U_{\mu}$
the 4-velocity, the extra force is given by \cite{Bertolami:2007gv}
\begin{equation}
f^{\mu}=\frac{1}{\rho+p}\left[\frac{\lambda F_{2}}{1+\lambda f_{2}}\left({\cal L_{M}}-p\right)\nabla_{\nu}R+\nabla_{\nu}p\right]h^{\mu\nu},\label{eq:extra-force}
\end{equation}
where $h_{\mu\nu}=g_{\mu\nu}+U_{\mu}U_{\nu}$ is the projection operator.

\section{TRAVERSABLE WORMHOLE GEOMETRIES SUPPORTED BY THE NONMINIMAL CURVATURE-MATTER
COUPLING}

\subsection{Wormhole Metric and the Gravitational Field Equations\label{sub:Wormhole-Metric-and}}

We consider the wormhole metric written as follows \cite{Morris:1988cz}:
\begin{equation}
ds^{2}=-e^{2\Phi(r)}dt^{2}+\frac{dr^{2}}{1-\frac{b(r)}{r}}+r^{2}\left(d\theta^{2}+\sin^{2}\theta d\phi^{2}\right),\label{eq:wormhole metric}
\end{equation}
where $\Phi(r)$ and $b(r)$ are arbitrary functions, usually referred
to as redshift and shape functions, respectively. The radial coordinate
has specific properties. Contrary to the proper length, $l$, which
is monotonic and that vanishes at the wormhole throat, the radial
coordinate is defined only in the interval $\left[r_{0},+\infty\right]$
where it is nonmonotonic with a minimum at the wormhole throat, $r_{0}$.
At this point we have a coordinate singularity: $b(r_{0})=r_{0}$.

Furthermore, functions $\Phi(r)$ and $b(r)$ must satisfy some additional
constrains \cite{Morris:1988cz}: $\left(1\right)$ The so-called
flaring out condition implies that at or close to the throat, $\left(b(r)-b'(r)r\right)/b^{2}(r)>0$.
This is the constraint that induces the violation of the NEC; $\left(2\right)$
Moreover, in order to have a proper length function that is finite
and well behaved, the condition $1-\frac{b(r)}{r}\geqslant0$ must
be satisfied everywhere; $\left(3\right)$ Finally, functions $\Phi(r)$
and $b(r)$ should also verify the condition $\left(r-b(r)\right)\Phi'(r)\rightarrow0$
as $r\rightarrow r_{0}$, which follows from the finiteness of the
energy density $\rho\left(r\right)$ and $b'(r)$.

These conditions, for functions $\Phi(r)$ and $b(r)$, ensure sensible
wormhole solutions. But if the goal is to obtain a traversable wormhole,
the existence of horizons must be prevented. Hence, the redshift function
$\Phi(r)$ must remain finite everywhere and should vanish as we approach
asymptotic flat regions. Additionally, there are a few quantitative
conditions that must be verified concerning the duration of the hypothetical
journey through the wormhole and about the forces felt by the hypothetical
traveler. These constraints are discussed in great detail in Ref.
\cite{Morris:1988cz}.

\subsection{Energy Conditions}

A wormhole must violate the NEC, and in GR this translates into the
condition $T_{\mu\nu}k^{\mu}k^{\nu}<0$ in the vicinity of the wormhole
throat. In a theory with nonminimal curvature-matter coupling, the
energy conditions were studied in Ref. \cite{Bertolami:2009cd} and
the condition to have wormhole solutions translates into $T_{\mu\nu}^{eff}k^{\mu}k^{\nu}<0$
as follow from Eq. (\ref{eq:Einstein Equation modified}). This is
a fundamental feature of our analysis since it allows, in principle,
for some values of the nonminimal coupling parameter $\lambda$, to
violate the NEC while satisfying for the stress-energy tensor of matter
the condition: $T_{\mu\nu}^{(m)}k^{\mu}k^{\nu}\geq0$. Furthermore,
wormhole solutions can be obtained even if matter satisfies the DEC.

\subsection{Time Machines}

Once a wormhole solution has been obtained it can be shown that one
can convert it into a time machine. For instance, following Ref. \cite{Morris:1988tu}
one way of doing such conversion consists in accelerating one of the
wormhole mouths close to the speed of light and then revert its motion
to its original location. This acceleration can be achieved by gravitational
or electromagnetic means. The metric that describes this procedure,
within the accelerated wormhole and outside but near its mouths, is
given by{\footnotesize 
\[
ds^{2}=-\left(1+g\left(t\right)lF\left(l\right)\cos\theta\right)e^{2\Phi}dt^{2}+dl^{2}+r^{2}\left(d\theta^{2}+\sin^{2}\theta d\phi^{2}\right),
\]
}where $l$ is the proper length, $\Phi$ is the same redshift function,
$F\left(l\right)$ is a form factor that localizes the acceleration
in one of the wormhole mouths and $g\left(t\right)$ is the acceleration
of that mouth as measured by its own asymptotic frame. Some other
alternative ways of producing time machines are also described in
Ref. \cite{Visser:1992tx}. In summary, the construction of a time
machine requires three indispensable steps: a stable traversable wormhole,
a time shift between the two mouths, and a pull to bring them close
together adiabatically.

The procedure of inducing a time-shift implies some additional conditions
on the type of acceleration applied to the wormhole mouth in order
to keep it traversable and stable. But the subtle point here is that
there is no additional constraints on the geometry. This means that
a stable traversable wormhole yields CTCs. As we shall see, in theories
with a nonminimal coupling between matter and curvature, stable configurations
that allow for time travel can be obtained even for ordinary matter,
that is, matter that satisfy the DEC.

\section{RESULTS}

\subsection{Specific case: $f_{1}(R)=f_{2}(R)=R$}

The field Eq. (\ref{eq:Einstein Equation modified}) are very complex,
and following Ref. \cite{Garcia:2010xb}, we consider the simplest
case of $f_{1}(R)=f_{2}(R)=R$ and introduce to start with the stress-energy
tensor of an anisotropic distribution of matter given by
\begin{equation}
T_{\mu\nu}=\left(\rho+p_{t}\right)U_{\mu}U_{\mu}+p_{t}g_{\mu\nu}+(p_{r}-p_{t})\chi_{\mu}\chi_{\nu}\label{eq:Anisotropic distribution of matter stress-energy tensor}
\end{equation}
where $U^{\mu}$ is the 4-velocity, $\chi^{\mu}$ is the unit spacelike
vector in the radial direction, i.e., $\chi^{\mu}=\sqrt{1-\frac{b(r)}{r}}\delta_{\ r}^{\mu}$,
$\rho(r)$ is the energy density, $p_{r}(r)$ is the radial pressure
measured in the direction of $\chi^{\mu}$ and $p_{t}(r)$ is the
tangential pressure measured in the orthogonal direction to $\chi^{\mu}$.

A relevant point is that there are various Lagrangian densities compatible
with the equation of state of a perfect fluid \cite{Bertolami:2008ab}.
Here we chose ${\cal L_{M}}=-\rho(r)$ \cite{Bertolami:2007vu}.

Having specified $f_{1}(R)\mbox{ and }f_{2}(R)$, Eq. (\ref{eq:Einstein Equation modified})
simplifies to
\begin{equation}
G_{\mu\nu}=(1+\lambda R)T_{\mu\nu}^{(m)}+2\lambda\left(\rho R_{\mu\nu}-\Delta_{\mu\nu}\rho\right)\label{eq:Simplified einstein field equation}
\end{equation}
where, for the wormhole metric Eq. (\ref{eq:wormhole metric}), the
Ricci scalar is given by

{\small 
\begin{equation}
R=\frac{2b'}{r^{2}}-2\left(1-\frac{b}{r}\right)\left[\Phi''+\frac{\Phi'}{r}\left(2-\frac{b'r-b}{2r(1-\frac{b}{r})}\right)+(\Phi')^{2}\right].\label{eq:Ricci scalar}
\end{equation}
}{\small \par}

Equation (\ref{eq:Simplified einstein field equation}) gives rise
to the following gravitational field equations:{\scriptsize 
\begin{equation}
\frac{b'}{r^{2}}+2\lambda\left(1-\frac{b}{r}\right)\left[\rho''+\frac{\rho'}{r^{2}}\left(2r-\frac{b'r-b}{2\left(1-\frac{b}{r}\right)}\right)\right]-\rho\left(1+\frac{2\lambda b'}{r^{2}}\right)=0,\label{eq:tt Einstein equation}
\end{equation}
}{\scriptsize \par}

{\scriptsize 
\[
p_{r}\lambda\left(\frac{2b'}{r^{2}}-2\left(1-\frac{b}{r}\right)\left[\Phi''+\frac{\Phi'}{r}\left(2-\frac{b'r-b}{2r\left(1-\frac{b}{r}\right)}\right)+\left(\Phi'\right)^{2}\right]\right)
\]
}{\scriptsize \par}

{\scriptsize 
\[
+2\lambda\rho\left(-\Phi''\left(1-\frac{b}{r}\right)+\frac{b'r-b}{2r^{2}}\Phi'-\left(1-\frac{b}{r}\right)\left(\Phi'\right)^{2}+\frac{b'r-b}{r^{3}}\right)
\]
\begin{equation}
+p_{r}+\frac{b}{r^{3}}+\left(1-\frac{b}{r}\right)\left(\frac{2\Phi'}{r}+2\lambda\rho'\left(\Phi'+\frac{2}{r}\right)\right)=0,\label{eq: rr Einstein equation}
\end{equation}
}{\scriptsize \par}

{\scriptsize 
\[
p_{t}r^{2}\lambda\left(\frac{2b'}{r^{2}}-2\left(1-\frac{b}{r}\right)\left[\Phi''+\frac{\Phi'}{r}\left(2-\frac{b'r-b}{2r\left(1-\frac{b}{r}\right)}\right)+\left(\Phi'\right)^{2}\right]\right)
\]
}{\scriptsize \par}

{\scriptsize 
\[
+2\lambda\rho\left(\frac{b'r+b}{2r}-r\Phi'\left(1-\frac{b}{r}\right)\right)+p_{r}r^{2}-\frac{b}{2r}+\frac{b'}{2}
\]
}{\scriptsize \par}

{\scriptsize 
\[
+2\lambda r^{2}\left(1-\frac{b}{r}\right)\left[\rho''-\rho'\left(\frac{b'r-b}{2r^{2}\left(1-\frac{b}{r}\right)}-\frac{1}{r}-\Phi'\right)\right]
\]
}{\scriptsize \par}

{\scriptsize 
\begin{equation}
-r^{2}\left(1-\frac{b}{r}\right)\left[\Phi''+\frac{\Phi'}{r}\left(1-\frac{b'r-b}{2r\left(1-\frac{b}{r}\right)}\right)+\left(\Phi'\right)^{2}\right]=0.\label{eq:theta theta einstein equation}
\end{equation}
}{\scriptsize \par}

That is, we have three Eqs. involving five unknown functions of $r$,
i.e., $\rho(r),p_{r}(r),p_{t}(r),b(r),\Phi(r)$. Thus, we have to
simplify our problem. An interesting possibility is to consider an
isotropic pressure ($p_{r}=p_{t})$ and specify a simple and plausible
energy density function $\rho(r)$ threading the wormhole. 

Notice that Eq. (\ref{eq:tt Einstein equation}), relating the functions
$b(r)$ and $\rho(r)$, can be integrated before any simplification:
\begin{equation}
b(r)=\left[\int\frac{re^{g(r)}\left(-\rho r+2\lambda\rho''r+4\lambda\rho'\right)}{\lambda\left(\rho'r+2\rho\right)-1}dr+C\right]\label{eq:b function depending on the energy density}
\end{equation}
where $C$ is an integration constant and $g(r)$ is a function defined
as
\begin{equation}
g(r)=\lambda\int\frac{3\rho'+2\rho''r}{\lambda\left(\rho'r+2\rho\right)-1}dr.\label{eq:auxiliary function for the b function}
\end{equation}

\subsection{Specific Solutions}

Following the procedure described above we consider for $p_{r}=p_{t}$
two different energy densities.

\subsubsection{Constant and localized energy density}

First, we examine the case of a constant energy density localized
within the region $r<r_{2}$ 
\begin{equation}
\rho_{1}(r)=\begin{cases}
\begin{array}{cc}
\rho_{0}, & r<r_{2}\\
0, & r>r_{2}
\end{array}\end{cases}\label{eq:energy density function}
\end{equation}
where $r_{2}$ is an arbitrary radial coordinate which we will fix
later in order to better determine our problem. 

With these conditions and neglecting any possible effects arising
from the discontinuity of the energy density at $r=r_{2}$, we obtain
the following shape function $b(r)$ from Eqs. (\ref{eq:b function depending on the energy density})
and (\ref{eq:auxiliary function for the b function}):
\begin{equation}
b_{1}(r)=\begin{cases}
\begin{array}{c}
Ar^{3}+C_{1},\\
C_{2},
\end{array} & \begin{array}{c}
r<r_{2}\\
r>r_{2}
\end{array}\end{cases}\label{eq:shape function}
\end{equation}
where $A=-\rho_{0}/3\left(2\rho_{0}\lambda-1\right)$  and $C_{1},C_{2}$
are integration constants. Imposing $b_{1}(r_{0})=r_{0}$, it allows
us to fix $C_{1}$ as 
\begin{equation}
C_{1}=r_{0}+\frac{\rho_{0}r_{0}^{3}}{3(2\lambda\rho_{0}-1)}.\label{eq:shape function integration constant}
\end{equation}
From the continuity at $r=r_{2}$ it follows that:

\begin{equation}
C_{2}=\frac{\rho_{0}}{3(2\lambda\rho_{0}-1)}\left(r_{0}^{3}-r_{2}^{3}\right)+r_{0}.\label{eq:shape function integration constant-1}
\end{equation}
 Moreover, we can set $C_{2}=0$ by a suitable choice of $r_{2}$.

Of course, the obtained shape function must satisfy the conditions
discussed in Sec. \ref{sub:Wormhole-Metric-and}. Therefore the parameters
of the theory are constrained by some inequalities. The shape function
Eq. (\ref{eq:shape function}) satisfies automatically all but the
flaring out condition. On its turn, the flaring condition implies
that:
\begin{equation}
\frac{\rho_{0}}{(2\lambda\rho_{0}-1)}>-\frac{1}{r_{0}^{2}}.\label{eq:condition on the parameters}
\end{equation}

\subsubsection{Exponentially decaying energy density}

The second case is an energy density given by
\begin{equation}
\rho_{2}(r)=\frac{\rho_{0}r_{0}}{r}e^{-\frac{r-r_{0}}{\sqrt{2\lambda}}},\label{eq:Energy Density-1}
\end{equation}
which satisfies the differential equation $-\rho_{2}r+2\lambda\rho_{2}''r+4\lambda\rho_{2}'=0$
that appears in Eq. (\ref{eq:b function depending on the energy density}).
The solution for $\rho_{2}(r)$ is real only if $\lambda\geq0$. This
choice for the energy density implies that the shape function is constant
and given by 
\begin{equation}
b_{2}(r)=r_{0}\label{eq:Shape Function-1}
\end{equation}
due to the condition $b_{2}(r_{0})=r_{0}$. 

The obtained shape function satisfies the conditions discussed in
Sec. \ref{sub:Wormhole-Metric-and}.

\subsection{Solutions for the redshift function $\Phi(r)$ and the pressure $p(r)$}

Using the matter distribution Eq. (\ref{eq:energy density function}),
the condition that $p_{r}=p_{t}$ and the solution Eq. (\ref{eq:shape function}),
we are left with Eqs. (\ref{eq: rr Einstein equation}) and (\ref{eq:theta theta einstein equation})
and two unknown functions ($\Phi(r)$, $p(r)$). Using these Eqs.
we can eliminate the pressure to obtain a nonlinear differential equation
for the redshift function $\Phi(r)$:
\[
\left(1-2\lambda\rho\right)\left[\frac{3b-b'r}{2r}-\Phi'\frac{b'r-b}{2}\right]
\]

\[
+r\left(1-\frac{b}{r}\right)\left(1-2\lambda\rho\right)\left(\Phi''+\left(\Phi'\right)^{2}+\frac{\Phi'}{r}\right)r
\]
\begin{equation}
+r\left(1-\frac{b}{r}\right)\left[2\Phi'+2\lambda\left(\rho'-\rho''r\right)\right]+\lambda\rho'\left(b'r-b\right)=0.\label{eq:diff eq for the redshift function-1}
\end{equation}

This is a very complex equation and an analytical solution to $\Phi(r)$
is out of reach. However, we are only interested in two limits: the
vicinity of the wormhole throat, where the violation of the NEC is
supposed to take place; and at infinity, where the solution is asymptotically
flat.

In the vicinity of $r=r_{0}$, $\left(1-\frac{b(r)}{r}\right)\rightarrow0$
and in this limit we obtain a simpler differential equation:
\begin{equation}
\left(1-2\lambda\rho\right)\left[\frac{3b-b'r}{2r}-\frac{b'r-b}{2}\Phi'\right]+\lambda\rho'\left(b'r-b\right)=0.\label{eq:Simplified diff equation for the redshift function-1}
\end{equation}

In the first case, using Eqs. (\ref{eq:energy density function})
and (\ref{eq:shape function}) and assuming that $\left(1-2\lambda\rho_{0}\right)\neq0$,
which has to be satisfied in order to achieve a well-defined shape
function, it follows that:
\begin{equation}
\Phi_{1}(r)=\log\left(\frac{C_{1}}{r^{3}}-2A\right)+C_{3},\label{eq:redshift function near r0}
\end{equation}
where $C_{3}$ is an integration constant.

Notice that the condition $\left(r-b_{1}\right)\Phi_{1}'\rightarrow0$
is satisfied as $r\rightarrow r_{0}$. Concerning the limit $r\rightarrow+\infty$,
$b_{1}(r)=\rho_{1}(r)=0$ by Eqs. (\ref{eq:energy density function}),
(\ref{eq:shape function}) and (\ref{eq:shape function integration constant-1}),
and we also expect that $r\left(\Phi_{1}'\right)^{2}\ll\Phi_{1}'$.
Hence, Eq. (\ref{eq:diff eq for the redshift function-1}) simplifies
to
\begin{equation}
r^{2}\Phi_{1}''+3r\Phi_{1}'=0,\label{eq:Simplified diff eq for the redshift function at infty-1}
\end{equation}
whose solution is 
\begin{equation}
\Phi_{1}(r)=-\frac{C_{4}}{2r^{2}}+C_{5},\label{eq:redshift function at infinity-1}
\end{equation}
where $C_{4},C_{5}$ are integration constants. Setting $C_{5}=0$,
we can easily see that $\Phi_{1}(r)\rightarrow0$ as $r\rightarrow+\infty$
as it should. We can also verify that the nonlinear terms are negligible
in this limit.

Substituting the solution for $\Phi_{1}(r)$ back into Eq. (\ref{eq: rr Einstein equation})
we obtain an algebraic equation for the pressure. The solution is
obtained following the same procedure. Close to the wormhole throat
we neglect the terms in $\left(1-\frac{b_{1}(r)}{r}\right)$ to obtain
\begin{equation}
p_{1}(r)=-\frac{Ar^{3}+C_{1}+2\lambda\rho_{0}\left[2Ar^{3}+\frac{C_{1}}{2}\right]}{\left[r^{2}+\lambda\left(6Ar^{3}+\frac{3C_{1}}{r}\right)\right]r}.\label{eq:pressure for r->r0-1}
\end{equation}

So that in the limit $r\rightarrow+\infty$, once again we can neglect
the nonlinear terms to obtain
\begin{equation}
p_{1}(r)=-\frac{2C_{4}}{r^{4}+4\lambda C_{4}},\label{eq:pressure for r->infinity-1}
\end{equation}
which vanishes for $r\rightarrow+\infty$.

Concerning the second energy density given by Eq. (\ref{eq:Energy Density-1}),
using Eq. (\ref{eq:Shape Function-1}), we have that in the vicinity
of $r=r_{0}$, $\left(1-\frac{b_{2}(r)}{r}\right)\rightarrow0$ and
in this limit the redshift is given by{\small 
\begin{equation}
\Phi_{2}(r)=-2\log\left(r\right)-\log\left(2\lambda\rho_{0}r_{0}e^{-\frac{r-r0}{\sqrt{2\lambda}}}-r\right)+C_{5}.\label{eq:Redshift Function, integral form-1}
\end{equation}
}where $C_{5}$ is an integration constant. In order to have a well-defined
redshift function one has to ensure that $2\lambda\rho_{0}r_{0}e^{-\frac{r-r0}{\sqrt{2\lambda}}}-r>0$
near the wormhole throat, which translates into the condition $2\lambda\rho_{0}>1$.

Notice that the condition $\left(r-b_{2}\right)\Phi_{2}'\rightarrow0$
is satisfied as $r\rightarrow r_{0}$. Concerning the limit $r\rightarrow+\infty$,
the energy density $\rho_{2}(r)$ decays very fast, hence, it can
be neglected along with its derivatives. Moreover, we can depreciate
the terms in $b_{2}/r$ in comparison to the unity and also $\left(\Phi_{2}'\left(r\right)\right)^{2}$
in comparison to $\Phi_{2}''\left(r\right)$. Thus, Eq. (\ref{eq:Simplified diff equation for the redshift function-1})
simplifies to
\begin{equation}
r^{2}\Phi_{2}''+3r\Phi_{2}'+\frac{3b_{2}}{r}=0,\label{eq:Simplified diff eq for the redshift function at infty-2}
\end{equation}
whose solution is 
\begin{equation}
\Phi_{2}(r)=\frac{3r_{0}}{2r}-\frac{C_{6}}{2r^{2}}+C_{7},\label{eq:redshift function at infinity-2}
\end{equation}
where $C_{5},C_{6}$ are integration constants. Setting $C_{7}=0$,
it can be easily seen that $\Phi(r)\rightarrow0$ as $r\rightarrow+\infty$
as it should. We can also verify that our considerations in neglecting
some terms are consistent.

Substituting the solution for $\Phi_{2}(r)$ back into Eq. (\ref{eq: rr Einstein equation})
leads to an algebraic equation for the pressure. The solution is obtained
following the same procedure. Close to the wormhole throat, we neglect
the terms in $\left(1-\frac{b_{2}(r)}{r}\right)$ to obtain
\begin{equation}
p_{2}(r)=\frac{\lambda\rho_{2}\left(r\Phi_{2}'+2\right)-1}{r(\frac{r^{2}}{b_{2}}-\lambda\Phi_{2}')}.\label{eq:pressure for r->r0-2}
\end{equation}

Once again, in the limit $r\rightarrow+\infty$ we can neglect the
energy density and its derivatives along with terms such as $b_{2}/r$
in comparison to unity to obtain
\begin{equation}
p_{2}(r)=-\frac{b_{2}+2\Phi_{2}'r^{2}}{r^{3}},\label{eq:pressure for r->infinity-2}
\end{equation}
which vanishes for $r\rightarrow+\infty$.

\subsection{Violation of the NEC}

Finally, we analyze the energy conditions of the obtained solutions.
This analysis consists in verifying if the violation of the NEC at
the vicinity of the wormhole throat, that is 
\begin{equation}
T_{\mu\nu}^{eff}k^{\mu}k^{\nu}<0\label{eq:NEC-1}
\end{equation}
with $k^{\mu}$ being a null vector. For simplicity, we choose $k^{\mu}$
to be radial. In the limit $r\rightarrow r_{0}$, where $\left(1-\frac{b(r)}{r}\right)\rightarrow0$,
the inequality Eq. (\ref{eq:NEC-1}) yields

{\footnotesize 
\begin{equation}
\left(\rho+p\right)\left(1+\frac{2\lambda b'}{r^{2}}\right)+\frac{\lambda}{r^{2}}\left(b'r-b\right)\left(\rho'+\frac{2\rho}{r}+\Phi'\left(\rho+p\right)\right)<0.\label{eq:NEC explicitly-1}
\end{equation}
}Restricting to the throat itself, at $r=r_{0}$, for the first case,
after using Eqs. (\ref{eq:shape function}), (\ref{eq:redshift function near r0})
and (\ref{eq:pressure for r->r0-1}), the NEC condition is equivalent
to
\begin{equation}
r_{0}^{2}\rho_{0}\left(1-\frac{2\lambda\rho_{0}}{2\lambda\rho_{0}-1}\right)<1.\label{eq:NEC inequality at r=00003Dr0}
\end{equation}
If the matter threading the wormhole satisfies $\rho_{0}>0$ , from\emph{
}Eq. (\ref{eq:NEC inequality at r=00003Dr0}) it follows for $\lambda$:

\begin{equation}
\lambda<\frac{1-\rho_{0}r_{0}^{2}}{2\rho_{0}}\ \mbox{or}\ \lambda>\frac{1}{2\rho_{0}}.\label{eq:NEC at r=00003Dr0 for solution 1}
\end{equation}
The first condition is incompatible with Eq. (\ref{eq:condition on the parameters}).
However, the second one is always compatible. Furthermore, from the
DEC, $\left|p(r_{0})\right|<\rho_{0}$, which, for $\lambda>1/2\rho_{0}$,
yields
\begin{equation}
\rho_{0}>\frac{1}{2\lambda}\left(1+\frac{r_{0}}{\sqrt{2\lambda+r_{0}^{2}}}\right).\label{eq:DEC at r=00003Dr0 for solution 1}
\end{equation}
Therefore, we conclude that wormhole solutions are obtained if $\lambda>1/2\rho_{0}$
and for ordinary matter if $\rho_{0}>\frac{1}{2\lambda}\left(1+\frac{r_{0}}{\sqrt{2\lambda+r_{0}^{2}}}\right)$.
However, we still have to require that $\Phi(r)$ ensues no horizons
and that $p(r)$ is well behaved everywhere. But we see that $b'(r)$
has a discontinuity at $r=r_{2}$ and both, the redshift function
and the pressure, depend on $b'(r)$. Therefore those quantities are
ill defined at this point and that gives rise to problems associated
to singularities, horizons, or unsuitable asymptotic behavior. Therefore,
this wormhole solution is not a traversable wormhole.

\begin{figure}
\centering{}\includegraphics[scale=0.6]{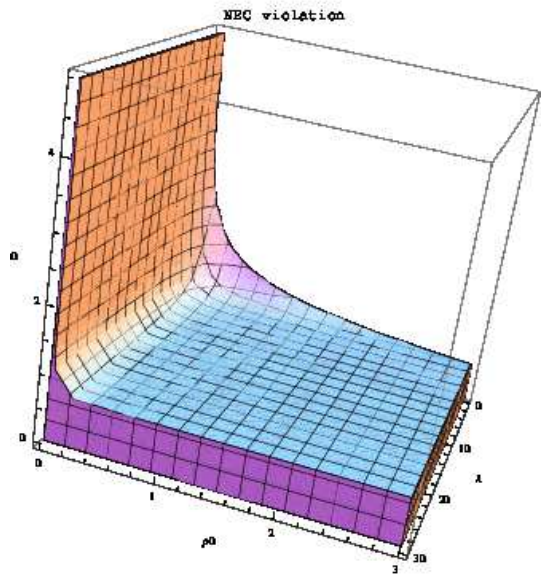}\caption{Region in the parameter space $\left(\lambda,\rho_{0},r_{0}\right)$
for which the NEC is violated at the wormhole throat. \label{fig:Region-in-the}}
\end{figure}

\begin{figure}
\begin{centering}
\includegraphics[scale=0.6]{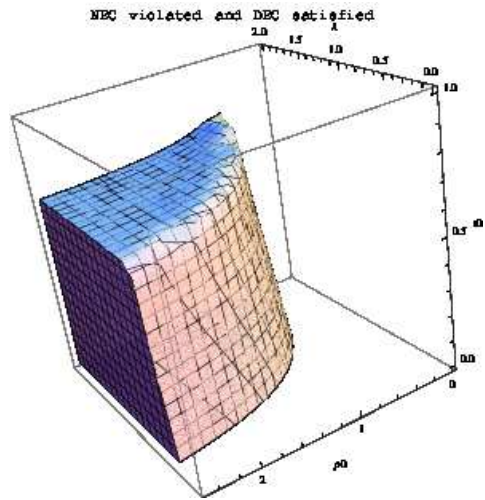}
\par\end{centering}

\caption{Region in the parameter space $\left(\lambda,\rho_{0},r_{0}\right)$
for which the NEC is violated and the DEC is satisfied at the wormhole
throat.\label{fig:DEC-NEC}}
\end{figure}

For the second case, the inequality Eq. (\ref{eq:NEC-1}) restricted
to the throat itself is satisfied by the 3D surface presented in Fig.
(\ref{fig:Region-in-the}).
 However, in order to transform this wormhole solution in a traversable
wormhole we have to impose two other constrains. First, we must ensure
that the redshift function Eq. (\ref{eq:Redshift Function, integral form-1})
is well defined, which means that $2\lambda\rho_{0}>1$. Moreover,
when we impose the constraint that the matter should verify at the
same time the DEC we obtain the region depicted in Fig. (\ref{fig:DEC-NEC}).

Therefore, we conclude from Fig. (\ref{fig:DEC-NEC}) that there are
regions in the parameter space $(\rho_{0},r_{0},\lambda)$ for which
traversable wormhole solutions with ordinary matter can be found.
The region close to $\lambda=0$ is not included in the solution space.
Because of the fact that the functions $\rho_{2}(r)$ and $b_{2}(r)$
are $C^{\infty}$ functions, the redshift function and the pressure
behave properly in the vicinity of the throat and at infinity, $\Phi(r)$
and $p(r)$ seem to be well behaved for $\lambda>0$. Thus, we conclude,
in opposition to the first studied energy density, that a matter distribution
as Eq. (\ref{eq:Energy Density-1}) presents no horizons and hence
the region depicted in Fig. (\ref{fig:DEC-NEC}) constitutes the space
of traversable wormhole solutions and therefore of time machines.

\section{DISCUSSION AND CONCLUSION}

GR admits a rich class of solutions such as wormholes and CTCs. Despite
the healthy skepticism about the existence and stability of these
solutions, the search of stable wormhole configurations and CTCs is
a topic of great interest. However, the construction of the traversable
wormholes and the formation of CTCs requires in GR the violation of
the NEC, which in turn demands the existence of exotic and yet unknown
forms of matter threading the wormhole. 

In this work, we have sought for traversable wormholes and CTCs solutions
in the context of $f(R)$ theories with nonminimal coupling between
curvature and matter. For simplicity the nonminimal coupling function
was chosen to be linear in the scalar curvature. There were studied
two different energy densities threading the wormhole: one constant
and localized within a certain region and another decaying and localized
near the wormhole throat. The field equations were then solved for
a perfect fluid. 

In the first case, the obtained solution for the shape function and,
in the limits $r\rightarrow r_{0}$ and $r\rightarrow\infty$, for
the redshift function and the pressure violate the NEC. This violation
ensures that the obtained solution is a wormhole, and it is verified,
at the wormhole throat for a positive energy density, provided the
coupling parameter of the theory satisfies the condition $\lambda>1/2\rho_{0}$.
Furthermore, if the energy density satisfies the inequality $\rho_{0}>\frac{1}{2\lambda}\left(1+\frac{r_{0}}{\sqrt{2\lambda+r_{0}^{2}}}\right)$
these wormhole solutions can be obtained even for ordinary matter.
Nevertheless, there is a discontinuity at an arbitrary scale of the
problem which is unavoidable and transforms the wormhole in a nontraversable
one. 

Concerning the energy density Eq. (\ref{eq:Energy Density-1}), the
obtained solution for the shape function and for the redshift function
violates the NEC if the parameters are within a region shown in Fig.
(\ref{fig:Region-in-the}). Therefore, this region ensures the existence
of wormhole solutions which can be created even with ordinary matter
if the parameters $\left(\rho_{0},r_{0}\right)$ and the coupling
parameter of the theory ($\lambda)$ are within the regions depicted
in Fig. (\ref{fig:DEC-NEC}). The key point is that in this second
case the found solutions are stable configurations and well behaved,
without horizons. So one can conclude that CTCs are, in this context,
unproblematic and allow for time travel if the quantitative conditions,
both for traversable wormholes and for the acceleration which produces
the time-shift, are satisfied.

Clearly, our solutions can be obtained if and only if $\lambda\neq0$
and $\lambda>0$, i.e. in the presence of the nonminimal coupling.
It is thus no surprise that the limit $\lambda\rightarrow0$ is out
of the solutions space. Of course, our solution reveals that the onus
of generating the wormhole solutions lies on the magnitude of the
nonminimal coupling for a given matter energy density (cf. condition
Eq. (\ref{eq:NEC at r=00003Dr0 for solution 1})). 
\begin{acknowledgments}
The work of one of us (O.B.) is partially supported by the FCT projects
PTDC/FIS/111362/2009 and CERN/FP/116358/2010.
\end{acknowledgments}

\end{document}